\documentclass[aps,prd,twocolumn,groupedaddress,showpacs]{revtex4}
\usepackage{epsf,epsfig,graphics,graphicx}
\begin{document}

\title{Collapse of vacuum bubbles in a vacuum}

\author{Kin-Wang Ng}
\email{nkw@phys.sinica.edu.tw}
\affiliation{Institute of Physics, Academia Sinica, Taipei,
Taiwan 11529, R.O.C.}
\author{Shang-Yung Wang}
\email{sywang@mail.tku.edu.tw}
\affiliation{Department of Physics, Tamkang University, Tamsui, Taiwan 25137, R.O.C.}

\vspace*{0.6 cm}
\date{\today}
\vspace*{1.2 cm}

\begin{abstract}
Motivated by the discovery of a plenitude of metastable vacua in a
string landscape and the possibility of rapid tunneling between
these vacua, we revisit the dynamics of a false vacuum bubble in a
background de Sitter spacetime. We find that there exists a large
parameter space that allows the bubble to collapse into a black
hole or to form a wormhole. This may have interesting implications
to inflationary physics.
\end{abstract}

\maketitle

\section{Introduction}

There has been a long interest in the study of the dynamics of
vacuum bubbles in different background
cosmology~\cite{Blau,berezin,aurilia,Nambu,khlopov,aguirre,simon}. Presumably,
these bubbles may be formed in the course of phase transitions in the early Universe
or arise from inhomogeneities generated by quantum fluctuations during inflation.
In these studies, the bubble is modelled as a spherically symmetrical
region of true or false vacuum which is separated by a domain wall from
the exterior world such as Minkowski, de Sitter, or Friedmann-Robertson-Walker spacetime.
The motion of the domain wall, which is analyzed under the
general-relativistic thin-wall approximation,
characterizes the global structure of the spacetime.
For example, dependent on initial conditions, an observer in the interior will describe
a inflating bubble universe while an observer in the exterior will describe the
bubble as a black hole.

Recently, the possibility of a string landscape~\cite{landscape} has opened
a new perspective on the vacuum bubble dynamics. The string landscape is considered as
a complicated potential energy function in a multidimensional moduli space
with a plentitude of local minima separated by barriers of various heights.
One of these minima corresponds to the vacuum where our Universe exists today
and the evolution of the very early Universe is determined by the route that
the Universe has taken in the landscape. However, a detailed mapping of the
structure of the landscape is still elusive.

In the string landscape, a local minimum is a metastable de Sitter vacuum with
certain energy scale, tunneling down successively through potential energy barriers
to vacua with lower vacuum energies via bubble nucleation. When tunneling is taking place,
various regions of the spacetime are inflating in various de Sitter vacua, resulting in
a very complicated global spacetime structure with a hierarchy of vacuum bubbles separated
by domain walls. Typically, the height of the barrier between the vacua is of string scale,
so the tunneling rate in the landscape is expected to be exponentially suppressed and far
below the expansion rate of the inflating region. Consequently, most part of the spacetime
undergo eternal inflation, with some regions occasionally tunneling to a landscape with a
flat potential that drives slow-roll inflation or to anti de Sitter vacua with negative
cosmological constant.

Development in the understanding of false vacuum decay in the string landscape
has recently been made. It was found that under certain conditions the decay rates
can be greatly enhanced even for tall barriers due to resonant tunneling
in the presence of multiple vacua~\cite{tye,sarangi,podolsky},
due to stringy effects such as tunneling via the Dirac-Born-Infeld action~\cite{brown,sarangi},
by an effect that facilitates giant leaps in the landscape~\cite{brown10},
or by having exponential prefactors~\cite{feng}. In light of this, it is possible that
the tunneling rate is not smaller than the expansion rate of the de Sitter
phase and thus a few or even copious vacuum bubbles may be formed within a Hubble radius.
This may lead to bubble collisions~\cite{kleban} with one of the domain
walls near our horizon, or the formation of black holes and wormholes that induce new metric
perturbation during slow-roll inflation~\cite{cho}.
Furthermore, sampling the multiple vacua with shallow barriers by quantum fluctuations
during inflation may populate domain walls that may affect the subsequent evolution of
the Universe~\cite{shiu}. All of these, though still speculative, may leave important
observable signatures of the string landscape.

In light of this, we reexamine the dynamics of a de Sitter bubble in a background
de Sitter spacetime. The bubble is assumed as spherically symmetric and separated
from the background by a thin shell. Then we use Israel's thin shell formalism~\cite{Israel}
to glue the bubble to the background, as was done in Refs.~\cite{Blau,berezin,aurilia,Nambu,aguirre}.
Although most of the global structure of the spacetime has been categorized,
the parameter space that we are interested here has not been fully explored.
The paper is organized as follows. In Sec.~\ref{shell}, we review
Israel's thin shell formalism. In Sec.~\ref{junction}, we derive the
equation that governs the motion of the shell and show that in general the
shell has peculiar velocity. In addition, we briefly review the
maximally extended coordinate systems of the de Sitter spacetime and
the Schwarzschild-de Sitter spacetime. The evolution
of the bubble is discussed in Sec.~\ref{bubble} and the numerical
results are summarized in Sec.~\ref{numer}.
Section~\ref{conc} is the conclusion.
Here we follow the sign convention and definitions in Ref.~\cite{MTW}.

\section{Thin Shell Formalism}\label{shell}

To describe the behavior of the shell which divides the
spacetime into two regions ${\cal M^+}$ and ${\cal M^-}$, it is simplest
to introduce a Gaussian normal coordinate system in the
neighborhood of the shell. If we denote by $\Sigma$ the
timelike (2+1)-dimensional spacetime hypersurface in which the shell
lies, we can introduce (2+1)-dimensional coordinates $x^i
\equiv (\tau, x^2, x^3)$ on $\Sigma$, where $i$ runs from $1$ to $3$.
Here $\tau $ is the proper-time variable that would be measured by an
observer comoving with the shell. In the Gaussian normal
coordinate system, a geodesic in a neighborhood of $\Sigma$
which is orthogonal to $\Sigma$ is taken as the third spatial
coordinate denoted by $\eta$. Thus, the full set of
coordinates is given by $x^{\mu} \equiv (x^i,\eta)$. We can
take $x^2 = \theta $ and $x^3 = \phi $ in the case of a
spherically symmetric shell. Then the metric can be written as
\begin{equation}
ds^2 = d{\eta}^2 + {\rm g}_{ij}dx^i dx^j.
\label{me1}
\end{equation}
We introduce an unit vector field ${\xi}^{\mu}(x)$ which is
normal to $\eta$ = const.\ hypersurfaces and pointing
from ${\cal M^-}$ to ${\cal M^+}$.
The induced metric on the
hypersurface $\Sigma$ can then be written as
\begin{equation}
h_{\mu \nu} = g_{\mu \nu} - {\xi}_{\mu} {\xi}_{\nu}.
\label{ind}
\end{equation}
In the Gaussian normal coordinates,
\begin{equation}
{\xi}^{\mu}(x) = {\xi}_{\mu}(x) = (0, 0, 0, 1)
\label{geo}
\end{equation}

The Gaussian normal coordinate system is suitable for the
Gauss-Codazzi formalism~\cite{MTW} in the neighborhood of $\Sigma$ with the
coordinate $\eta$ orthogonal to the slices near $\Sigma$. The
extrinsic curvature is defined by
\begin{equation}
K_{ij} ={\xi}_{i;j}~,
\end{equation}
or equivalently
\begin{equation}
K_{ij} = -{\Gamma}^{\eta}_{ij} = {1 \over 2}
{\partial}_{\eta}{\rm g}_{ij}~.
\label{ext}
\end{equation}
It is evident that $K_{ij}$ is a symmetric three-tensor.

The Einstein equations are
\begin{equation}
R_{\mu \nu} - \frac{1}{2} g_{\mu \nu} R = 8 \pi G T_{\mu \nu}~,
\label{Einstein}
\end{equation}
where $R_{\mu \nu}$ is the Ricci tensor,
$R$ is the Ricci scalar, and $T_{\mu \nu}$
is the matter energy-momentum tensor.
In addition, the energy-momentum conservation law implies
\begin{equation}
{T^{\mu\nu}}_{;\nu}=0~,
\end{equation}
where the semicolon represents the four-dimensional covariant derivative.
Assuming that the energy-momentum tensor only
has $\delta$-function singularity on the shell, one can rewrite it
in terms of Gaussian normal coordinates as
\begin{equation}
T_{\mu \nu}= S_{\mu\nu}(x^i)\delta(\eta) + (\mbox{\rm regular terms})~,
\end{equation}
where $S_{\mu\nu}$ is the surface stress-energy tensor of the shell.
Adopting the thin shell formalism~\cite{Israel},
we can rewrite the Einstein equations and the
energy-momentum conservation law as
\begin{equation}\label{K}
[{K_j}^i] = - 8\pi G ({S_j}^i - {1 \over 2} {\delta}_j^i S)~,
\label{jun}
\end{equation}
\begin{equation}\label{S}
{S_i}^j_{|j}+[{T_i}^\eta]=0~,
\end{equation}
and
\begin{equation}\label{SK}
\{{K_j}^i\} {S_j}^i - [{T_\eta}^\eta]=0~,
\end{equation}
where $\{K_{ij}\} = {1 \over 2}( K_{ij}^+ + K_{ij}^-)$,
 $[A] = {A}^+ - {A}^-$ for any quantity $A$,
and a subscript vertical bar demotes the
three-dimensional covariant derivative.

By the symmetry analysis and the energy-momentum conservation
law~\cite{berezin}, we have
\begin{equation}\label{emS}
S^{\mu\nu}=\sigma(\tau)U^{\mu}U^{\nu}-\zeta(\tau)(h^{\mu\nu}-
U^{\mu}U^{\nu})~,
\end{equation}
where $U^{\mu}=(1,0,0,0)$ is the four-velocity of the shell in
the Gaussian normal coordinates. Here $\sigma$ and $\zeta$ are
the surface energy density and the surface tension of the shell,
respectively.
In the case of a spherically symmetric shell,
the induced metric on the shell can be written as
\begin{equation}
{ds^2|}_{\Sigma}=-d\tau^2+r^2(\tau)d\Omega^{2}~,
\end{equation}
where $r(\tau)$ is the proper circumferential radius of the shell.
Eq.~(\ref{S}) then reduces to
\begin{equation}\label{sigma}
\dot{\sigma}=-2(\sigma-\zeta)\frac{\dot{r}}{r}~,
\end{equation}
where the overdot denotes a derivative with respect to $\tau$.
Using the thin-shell approximation and the underlying
field-theoretical dynamics of the scalar field which comprises the
shell~\cite{Blau}, one can show that $\zeta=\sigma$. From
Eq.~(\ref{sigma}), it follows that $\dot{\sigma}=0$.
Thus, Eq.~(\ref{emS}) reduces
to
\begin{equation}\label{emS2}
S^{\mu\nu}=-\sigma h^{\mu\nu}~,
\end{equation}
where $\sigma$ is a constant independent of $\tau$.
By combining Eq.~(\ref{K}) with Eq.~(\ref{emS2}), one finds
\begin{equation}\label{EOM:K}
{K_j}^i({\cal M^+})-{K_j}^i({\cal M^-})=-4\pi\sigma G
{\delta_j}^i~.
\end{equation}

\section{Junction Conditions}\label{junction}

For the system under consideration, we have
\begin{equation}
T_{\mu \nu}=\left\{
\begin{array}{ll}
-(\Lambda_{1}/8\pi G)g_{\mu\nu}&\mbox{interior of the shell},\\
-(\Lambda_{2}/8\pi G)g_{\mu\nu}&\mbox{exterior of the shell},
\label{EMtensor}
\end{array}
\right.
\end{equation}
where $\Lambda_1$ and $\Lambda_2$ are the cosmological constants of the shell
interior and exterior, respectively. For convenience, we will consider
positive values for both $\Lambda_1$ and $\Lambda_2$ in the following.
At the end, we will extend the results for negative $\Lambda_1$ and $\Lambda_2$.

>From Birkhoff theorem, we can easily write down the most general
$O(3)$-symmetric solutions to the Einstein equations for a region of
spacetime with a non-vanishing cosmological constant. They are the
de Sitter~(dS) and Schwarzschild-de Sitter~(SdS) spacetimes. Thus,
the metrics of the interior and exterior spacetime in the static
coordinate system are
\begin{equation}
ds^{2}_{\pm}=-F_{\pm}(r)dt^2+{F_{\pm}}^{-1}(r)dr^2 +
r^2 d\Omega^2~,
\end{equation}
where
\begin{equation}
F_{\pm}(r)=\left\{
\begin{array}{ll}
1-{\chi_1}^{2}r^{2}&\mbox{interior spacetime},\\
1-2MG/r-{\chi_2}^{2}r^{2}&\mbox{exterior spacetime}.
\end{array}
\right.
\end{equation}
Here $M$ is an as yet undetermined parameter,
${\chi_1}^{2}=\Lambda_{1}/3$,
${\chi_2}^{2}=\Lambda_{2}/3$, and
$d\Omega^2=d\theta^2+\sin^2\theta d\phi^2$.
The corresponding $\theta$-$\theta$ components of the
extrinsic curvatures of $\Sigma$, as it is
embedded into the de Sitter and Schwarzschild-de Sitter spacetimes,
can be easily calculated in the static coordinates
by a standard recipe~\cite{Israel,Blau,berezin}.
They are
\begin{equation}\label{KdS}
K_{\theta\theta}(\mbox{\rm dS})=r\beta_{\rm dS}
\end{equation}
and
\begin{equation}\label{KSdS}
K_{\theta\theta}(\mbox{\rm SdS})=r\beta_{\rm SdS}~,
\end{equation}
where
\begin{equation}\label{beta:dS}
\beta_{\rm dS}=\pm(1-{\chi_1}^2 r^2+\dot{r}^2)^{1/2}
\end{equation}
and
\begin{equation}\label{beta:SdS}
\beta_{\rm SdS}=\pm(1-2MG/r-{\chi_2}^2 r^2+\dot{r}^2)^{1/2}~.
\end{equation}
Substituting Eqs.~(\ref{KdS}) and (\ref{KSdS}) into Eq.~(\ref{EOM:K}),
we get the equation of motion of the shell,
\begin{equation}\label{EOM:r}
\dot{r}^2=\frac{{\chi_+}^4 r^2}{4 \kappa^2}+\frac{MG(\kappa^2+
{\chi_2}^2-{\chi_1}^2)}{\kappa^2 r}+\frac{M^2 G^2}{\kappa^2 r^4}-1~,
\end{equation}
where
\begin{equation}
\kappa\equiv 4\pi G \sigma \ge 0
\end{equation}
and
\begin{equation}\label{chi+}
{\chi_+}^2\equiv\left[(\kappa^2+{\chi_1}^2-{\chi_2}^2)^2
      +(2\kappa\chi_2)^2\right]^{1/2}~.
\end{equation}
It is perhaps impossible to analytically solve Eq.~(\ref{EOM:r}).
However,
the asymptotic behavior of $r$ is easily obtained. We have
\begin{equation}\label{r:infinity}
r(\tau) \propto \exp({\chi_+}^2 \tau/2 \kappa)~~~~~~~\mbox{\rm as $r \rightarrow \infty$,}
\end{equation}
and
\begin{equation}
r(\tau) \propto \left( \frac{3 M G \tau}{\kappa} \right)^{1/3}~~~~
\mbox{\rm as $r \rightarrow 0$.}
\end{equation}

Now we are going to show that the shell is not static relative to an observer
using the comoving coordinates.
Since the Schwarzschild-de Sitter spacetime approaches the de Sitter
spacetime asymptotically, it serves our purpose by simply using
its asymptotic form. For a de Sitter spacetime, the transformation
between the static coordinates $(t,r)$ and the comoving coordinates
$(\bar{t},\bar{r})$ is well-known, given by~\cite{birrell}
\begin{equation}
\left\{
\begin{array}{l}
\bar{t} = t + \frac{1}{2\chi}\ln |1-\chi^2 r^2|~,\\
\bar{r} = r e^{-\chi t} |1-\chi^2 r^2|^{-1/2}~,
\end{array}
\right.
\end{equation}
where we have temporarily suppressed the subscript of $\chi$ for convenience.
>From the viewpoint of an exterior observer in the asymptotic region,
the proper circumferential radius $r$ is related to the comoving
radius $\bar{r}$ by
\begin{equation}
\bar{r} = \frac{r}{a(\bar{t})}~,
\end{equation}
where $a(\bar{t})=\exp(\chi_2 \bar{t})$ is the scale factor of the exterior
spacetime. The proper-time rate of change of $\bar{r}$ is
\begin{equation}
\left(\frac{d\bar{r}}{d\tau}\right)_{\rm SdS} = \frac{\dot{r}-\chi_2 r
(1-{\chi_2}^2 r^2 + \dot{r}^2)^{1/2}}{a |1-{\chi_2}^2 r^2|}~,
\end{equation}
where $a$ is now given as a function of $\tau$.
Compared with Eq.~(\ref{EOM:r}), we find that in general
$d\bar{r}/d\tau \ne 0$, i.e., the shell cannot be
evolving with a fixed comoving scale.
Similarly, the proper-time rate of change of $\bar{t}$ is
\begin{equation}
\left( \frac{d\bar{t}}{d\tau}\right)_{\rm SdS}
 = \frac{(1-{\chi_2}^2 r^2 + \dot{r}^2)^{1/2} -
\chi_2 r \dot{r}}{|1-{\chi_2}^2 r^2|}~.
\end{equation}
Using $d\bar{r}/d\bar{t}=(d\bar{r}/d\tau)/(d\bar{t}/d\tau)$
and Eq.~(\ref{EOM:r}), we have
\begin{equation}\label{v:SdS}
\left(\frac{d\bar{r}}{d\bar{t}}\right)_{\rm SdS}
\rightarrow a^{-1}(\bar{t})\left[
1+ \left(\frac{2 \kappa \chi_2}{\kappa^2 +{\chi_1}^2 -{\chi_2}^2}
\right)^{2}\right]^{-1/2}~,
\end{equation}
as $r\rightarrow\infty$.
On the other hand,
from the viewpoint of an interior observer,
we can repeat the above arguments and obtain
\begin{equation}\label{v:dS}
\left(\frac{d\bar{r}}{d\bar{t}}\right)_{\rm dS}
\rightarrow a^{-1}(\bar{t})\left[
1+ \left(\frac{2 \kappa \chi_1}{\kappa^2 -{\chi_1}^2 +{\chi_2}^2}
\right)^{2}\right]^{-1/2}~,
\end{equation}
as $r\rightarrow\infty$.
Hence, the necessary condition for a shell moving with a fixed comoving
scale is
\begin{equation}
\kappa^2=|{\chi_1}^2-{\chi_2}^2|.
\end{equation}
This means that the surface energy density of the shell has to compensate
the difference of the vacuum energy densities across the shell.
Moreover, the shell cannot be comoving with respect to both observers
simultaneously.
Note that if $\kappa=0$, the shell will become null.

Before solving the evolution of the bubble, we should address some
issues regarding the topological structure of the spacetime.
Firstly, there is a sign ambiguity in the expression of the
extrinsic curvature~(\ref{beta:dS}) and in Eq.~(\ref{beta:SdS}) as well.
Moreover, in the static coordinate system,
there exist coordinate singularities at the event horizon,
which are indeed caused by a poor choice of coordinates.
Nevertheless, the sign of the extrinsic curvature
and the physical interpretation are manifest when it is expressed in terms
of the Kruskal-like maximally extended coordinates.

The Gibbons-Hawking maximally extended coordinate system
$(v,u,\theta,\phi)$
for the de-Sitter spacetime is well known~\cite{Gibbons,Blau},
and it is related to the static coordinate system for $r<{\chi}^{-1}$ by
\begin{eqnarray}
u&=&
\left[\frac{1-\chi r}{1+\chi r}\right]^{1/2} \cosh(\chi t)~,\nonumber\\
v&=&
\left[\frac{1-\chi r}{1+\chi r}\right]^{1/2} \sinh(\chi t)~.\nonumber
\end{eqnarray}
The new metric is given by
\begin{equation}\label{dS:met}
ds^2=\chi^{-2}(1+\chi r)^2(-dv^2+du^2)+r^2 d\Omega^2~,
\end{equation}
where $r$ is a function of $u$ and $v$ given by
\begin{equation}
\frac{1-\chi r}{1+\chi r} = u^2 -v^2~.
\end{equation}
The metric~(\ref{dS:met}) can be maximally extended over the
entire $u$-$v$ plane, subject to the constraint
$|u^2-v^2|<1~$.
The coordinate system then covers the entire de Sitter spacetime.
In terms of the Gibbons-Hawking maximally extended coordinate system,
we can rewrite $\beta_{\rm dS}$ as
\begin{equation}
\beta_{\rm dS}=-{\chi_1}^{-1}(1+\chi_1 r)^2(u\dot{v}-v\dot{u})~.
\end{equation}

In the case of the Schwarzschild-de Sitter spacetime,
the maximally extended coordinates
was first derived by Bazanski and Ferrari~\cite{Bazanski}.
Unlike the Schwarzschild or the de Sitter spacetime,
the entire Schwarzschild-de Sitter spacetime
cannot be covered by a single
coordinate patch. This arises due to the fact that there exist
both black-hole and cosmological event horizons.

Recall that the Schwarzschild-de Sitter metric is given by
\begin{equation}
ds^{2}=-F(r)dt^2+F^{-1}(r)dr^2 +
r^2 d\Omega^2~,
\end{equation}
where
\begin{equation}
F(r)=1-\frac{2GM}{r}-\chi^2 r^2~.
\end{equation}
The location of the horizon in the Schwarzschild-de Sitter spacetime
can be found by solving the equation
\begin{equation}\label{rH}
 F(r)=0~.
\end{equation}
We define an extremal mass by $M_{\rm ext}\equiv 1/(\sqrt{27}\chi G)$,
which turns out to be a relevant mass scale in our analysis.
When $M>M_{\rm ext}$, Eq.~(\ref{rH}) has no positive root and hence
there does not exist any horizon. In the case of $0<M < M_{\rm ext}$,
there are two positive roots, $r_{\rm H}$ and $r_{\rm C}$
($r_{\rm H}< r_{\rm C}$), which satisfy
\begin{equation}
{r_{\rm H}}^2+r_{\rm H}r_{\rm C}+{r_{\rm C}}^2=\chi^{-2}
\end{equation}
and
\begin{equation}
r_{\rm H}r_{\rm C}(r_{\rm H}+r_{\rm C})=2M\chi^{-2}~.
\end{equation}
When $M = M_{\rm ext}$, two positive roots become identical.
We define the quantities $\delta_{\rm H}$ and $\delta_{\rm C}$
for future use by
\begin{equation}
\delta_{\rm H}=\frac{r_{\rm H}}{1-3 \chi^{2} r_{\rm H}^2}~,\hspace{1cm}
\delta_{\rm C}=-\frac{r_{\rm C}}{1-3 \chi^{2} r_{\rm C}^2}~.
\end{equation}
The coordinate patch which can be maximally extended to cover
the entire neighborhood of the black-hole horizon is given by
\begin{eqnarray}
u&=&\left[\frac{(r-r_{\rm H})(r+r_{\rm H}+r_{\rm C})^{x-1}}
{(r_{\rm C}-r)^x}\right]^{\frac{1}{2}}\cosh\left[
\frac{t}{2\delta_{\rm H}}\right]~,\nonumber\\
v&=&\left[\frac{(r-r_{\rm H})(r+r_{\rm H}+r_{\rm C})^{x-1}}
{(r_{\rm C}-r)^x}\right]^{\frac{1}{2}}\sinh\left[
\frac{t}{2\delta_{\rm H}}\right]~,\nonumber
\end{eqnarray}
where $r_{\rm H}\le r<r_{\rm C}$
and $x=\delta_{\rm C}/\delta_{\rm H}$.
The new metric is given by
\begin{eqnarray}\label{SdS:met1}
ds^2&=&4\chi^2\delta_{\rm H}^2(r_{\rm C}-r)^{1+x}
(r+r_{\rm H}+r_{\rm C})^{2-x}(-dv^2+du^2)\nonumber\\
&&+r^{2}d\Omega^2~,
\end{eqnarray}
where $r$ is a function of $u$ and $v$ given by
\begin{equation}
\frac{(r-r_{\rm H})(r+r_{\rm H}+r_{\rm C})^{x-1}}
{(r_{\rm C}-r)^x}=u^2-v^2~.
\end{equation}
The metric~(\ref{SdS:met1}) can be maximally extended over the
entire $u$-$v$ plane, subject to the constraint
\begin{equation}
u^2-v^2> - r_{\rm H}r_{\rm C}^{-x}(r_{\rm H}+r_{\rm C})^{x-1}~.
\end{equation}
The coordinate system then covers the entire
neighborhood of the black-hole horizon in the
Schwarzschild-de Sitter spacetime.

Similarly, the coordinate patch
which can be maximally extended to cover
the entire neighborhood of the
cosmological horizon is given by
\begin{eqnarray}
u&=&\left[\frac{(r_{\rm C}-r)}
{(r-r_{\rm H})^{\frac{1}{x}}
(r+r_{\rm H}+r_{\rm C})^{1- \frac{1}{x}}}\right]^
{\frac{1}{2}}
\cosh\left[\frac{t}{2\delta_{\rm C}}\right]~,\nonumber\\
v&=&\left[\frac{(r_{\rm C}-r)}{(r-r_{\rm H})^{\frac{1}{x}}(r+r_{\rm H}+r_{\rm C})^{1-\frac{1}{x}}}
\right]^{\frac{1}{2}}
\sinh\left[\frac{t}{2\delta_{\rm C}}\right]~,\nonumber
\end{eqnarray}
for $r_{\rm H}< r \le r_{\rm C}$.
The new metric is given by
\begin{eqnarray}\label{SdS:met2}
ds^2&=&4\chi^2\delta_{\rm C}^2(r-r_{\rm H})^{1+\frac{1}{x}}
(r+r_{\rm H}+r_{\rm C})^{2-\frac{1}{x}}(-dv^2+du^2)\nonumber\\
&&+r^{2}d\Omega^2~,
\end{eqnarray}
where $r$ is a function of $u$ and $v$ given by
\begin{equation}
\frac{r_{\rm C}-r}{(r-r_{\rm H})^{\frac{1}{x}}
(r+r_{\rm H}+r_{\rm C})^{1-\frac{1}{x}}}
=u^2-v^2~.
\end{equation}
The metric~(\ref{SdS:met2}) can be maximally extended over the
entire $u$-$v$ plane, subject to the constraint
\begin{equation}
u^2-v^2> - 1~.
\end{equation}
The coordinate system then covers the entire
neighborhood of the cosmological horizon in the
Schwarzschild-de Sitter spacetime.

In terms of the Bazanski-Ferrari
maximally extended coordinate system discussed
above, we can rewrite $\beta_{\rm SdS}$ as
\begin{equation}
\beta_{\rm SdS}=\left\{
\begin{array}{l}
2{\chi_2}^2\delta_{\rm H}(r_{\rm C}-r)^{1+x}(r+r_{\rm H}+r_{\rm C})^{2-x}
(u\dot{v}-v\dot{u})\\
~~~~~~~~~~~~~~~~~\mbox{for black hole horizon,}\\[2pt]
-2{\chi_2}^2\delta_{\rm C}(r-r_{\rm H})^{1+\frac{1}{x}}
(r+r_{\rm H}+r_{\rm C})^{2-\frac{1}{x}}
(u\dot{v}-v\dot{u})\\
~~~~~~~~~~~~~~~~~\mbox{for cosmological horizon.}
\end{array}
\right.
\end{equation}

\section{Evolution of the Bubble}\label{bubble}

To discuss the evolution of the bubble and the corresponding spacetime
structure,
we now introduce the dimensionless variables
\begin{equation}
z^3\equiv \frac{r^{3}}{r_{0}^{3}}~,
\end{equation}
and
\begin{equation}
\tau^{\prime}\equiv \frac{{\chi_+}^2\tau}{2\kappa}~,
\end{equation}
where
\begin{equation}
r_{0}^3\equiv\frac{2 M G}{{\chi_+}^2}~.
\end{equation}
Eq.~(\ref{EOM:r}) can be rewritten as~\cite{Blau,berezin,aurilia,Nambu}
\begin{equation}
   \left(\frac{dz}{d\tau^{\prime}}\right)^2+V(z)=E,
\end{equation}
 where
\begin{equation}
   V(z)=-\left(z-\frac{1}{z^2}\right)^2-\frac{\gamma^2}{z}~,
\label{Vz}
\end{equation}
\begin{equation}
   E=-\frac{4\kappa^2}{(2GM)^{2/3}{\chi_+}^{8/3}}~,
\label{Mparameter}
\end{equation}
and
\begin{equation}
\gamma^2=2+2(\kappa^2+{\chi_2}^2-{\chi_1}^2)/{\chi_+}^2~.
\end{equation}
Note that $0\le\gamma^2\le 4$.
The equation of motion is equivalent to that of a classical particle moving in
one dimension subject to the potential $V(z)$.
The asymptotic behaviors of the potential $V(z)$ are $V(z) \sim -1/z^4$
for small $z$ and $V(z) \sim -z^2$ for large $z$. One can show that
$d^2 V/d z^2 < 0$ for all $z$ and $V(z)$ has one maximum at $z_m$, where
\begin{equation}
{z_m}^3=\frac{1}{2}\{[8+(1-\gamma^2/2)^2]^{1/2}-(1-\gamma^2/2)\}~.
\end{equation}
The maximum value of the potential is given by
\begin{equation}
V(z_m)= -\frac{3({z_m}^6-1)}{{z_m}^4}~.
\end{equation}
In terms of the new variables, we can also rewrite $\beta_{\rm dS}$
and $\beta_{\rm SdS}$ as
\begin{eqnarray}
\beta_{\rm dS}&=&\left(\frac{GM}{r_0^2\kappa}\right)\frac{1}{z^2}
\left[1-\left(\frac{{\chi_1}^2
 -{\chi_2}^2-\kappa^2}{{\chi_+}^2}\right)z^3\right]~,\\
\beta_{\rm SdS}&=&\left(\frac{GM}{r_0^2\kappa}\right)\frac{1}{z^2}
\left[1-\left(\frac{\kappa^2+{\chi_1}^2
 -{\chi_2}^2}{{\chi_+}^2}\right)z^3\right]~.
\end{eqnarray}
If $\kappa^2 < {\chi_1}^2 - {\chi_2}^2$, then $\beta_{\rm dS}$
will change sign at $z=z_{\rm dS}$, where
\begin{equation}
z_{\rm dS}=\left( \frac{{\chi_+}^2}{{\chi_1}^2
 -{\chi_2}^2-\kappa^2} \right)^{1/3}~.
\end{equation}
If $\kappa^2 > {\chi_2}^2 - {\chi_1}^2$, then $\beta_{\rm SdS}$
will change sign at $z=z_{\rm SdS}$, where
\begin{equation}
z_{\rm SdS}=\left( \frac{{\chi_+}^2}{\kappa^2+{\chi_1}^2
 -{\chi_2}^2} \right)^{1/3}~.
\end{equation}
In case of both $\beta_{\rm dS}$ and $\beta_{\rm SdS}$ changing sign, since
\begin{equation}
\frac{z_{\rm dS}}{z_{\rm SdS}}=\left(
1+\frac{2 \kappa^2}{{\chi_1}^2-{\chi_2}^2-\kappa^2}\right)^
{1/3} \ge 1~,
\end{equation}
it follows that $z_{\rm dS}\ge z_{\rm SdS}$ and the equality holds for
$\kappa=0$.

The location of the horizon in the Schwarzschild-de Sitter spacetime is
determined by the equation,
\begin{equation}
1-\frac{2GM}{r}-{\chi_2}^2 r^2=0,
\end{equation}
and in terms of the dimensionless variables, it can be written as
\begin{eqnarray}
E&=&-\left(\frac{4\kappa^2}{{\chi_+}^2}\right)\frac{1}{z}-
           \left(\frac{4\kappa^2{\chi_2}^2}{{\chi_+}^4}\right)z^2 \nonumber\\
&=&V(z)+\frac{1}{z^4}
\left[1-\left(\frac{\kappa^2+{\chi_1}^2
       -{\chi_2}^2}{{\chi_+}^2}\right)z^3\right]^2~.
\label{outhorizon}
\end{eqnarray}
This curve is tangent to $V(z)$ at $z=z_{\rm SdS}$.
Similarly, the location of the horizon in the inner de Sitter spacetime is given
by the equation, $1-{\chi_1}^2 r^2=0$, which can be written as
\begin{eqnarray}
E&=&-\left( \frac{4\kappa^2 {\chi_1}^2}{{\chi_+}^4} \right) z^2 \nonumber\\
&=&V(z)+\frac{1}{z^4}\left[1-
\left(\frac{{\chi_1}^2
       -{\chi_2}^2-\kappa^2}{{\chi_+}^2}\right)z^3\right]^2.~
\label{inhorizon}
\end{eqnarray}
This curve is tangent to $V(z)$ at $z=z_{\rm dS}$.

\section{Numerical Results}\label{numer}

The system has four mass scales: $\chi_1$, $\chi_2$, $\kappa$, and $M$, which are
treated as free parameters in the present consideration. In the context of string landscape,
$\chi_1$ and $\chi_2$ are given by the cosmological constants of the local minima, and
$\kappa$ can be calculated as long as the profile of the
potential barrier between the minima is known. The initial conditions of the nucleated bubble
determine the value of $M$. So far, we have assumed positive cosmological constants.
For negative ones, the derivations in the previous section still hold after making the
change: $\chi^2\rightarrow -\chi^2$. A general discussion of the system has been given
in Ref.~\cite{aurilia}, but there lacks details about the parameter space that we are
concerned here. In Ref.~\cite{aguirre}, the authors considered $\chi_1>\chi_2>0$ only.
In Ref.~\cite{Nambu}, only a small $\kappa$ was considered in the context of quantum
fluctuations during inflation. In the following, we will follow the methodology in
Ref.~\cite{Nambu} that is suited to our purposes.

\begin{figure}
\leavevmode
\hbox{
\epsfxsize=3in
\epsffile{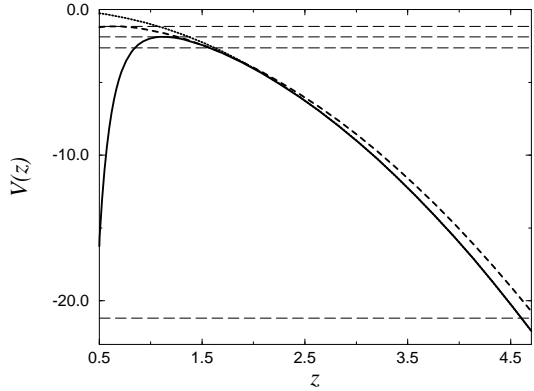}}
\caption{Graph of $V(z)$ (solid line), de Sitter (dashed line)
and Schwarzschild-de Sitter (dotted line)
horizon lines for $\chi_1=3.0$, $\chi_2=2.9$, and $\kappa=0.7375$
(in unit of a mass scale).
Long-dashed horizontal lines denote the values of $E$ corresponding to
$M_{\rm ext}$, $M_{\rm max}$, $M_{\rm SdS}$, and $M_{\rm dS}$, respectively.}
\label{figure1}
\end{figure}

\begin{figure}
\leavevmode
\hbox{
\epsfxsize=3in
\epsffile{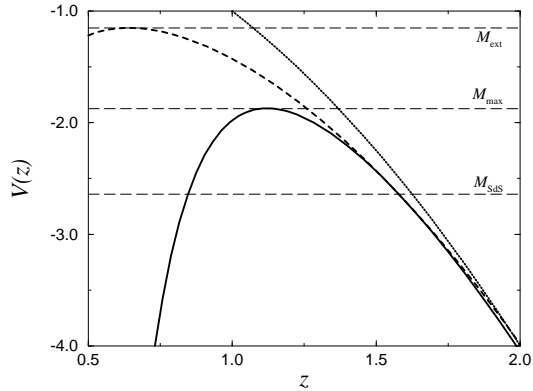}}
\caption{Blow-up graph of the upper-left region around the peak of
the potential in Fig.~1.}
\label{figure2}
\end{figure}

The plots of the curves of $V(z)$ in Eq.~(\ref{Vz}) and the event horizons
in Eqs.~(\ref{outhorizon}) and (\ref{inhorizon}) are shown in Fig.~\ref{figure1}
and Fig.~\ref{figure2}, for a representative case
using $\chi_1=3.0$, $\chi_2=2.9$, and $\kappa=0.7375$,
in unit of an arbitrary mass scale. Also drawn are four horizontal lines each of
which denotes the value of $E$ corresponding to the Schwarzschild mass $M$ given
in Eq.~(\ref{Mparameter}).
$M_{\rm ext}$ is the extremal mass, $M_{\rm max}=V(z_m)$,
$M_{\rm SdS}=V(z_{\rm SdS})$, and $M_{\rm dS}=V(z_{\rm dS})$.

\begin{figure}
\leavevmode
\hbox{
\epsfxsize=3in
\epsffile{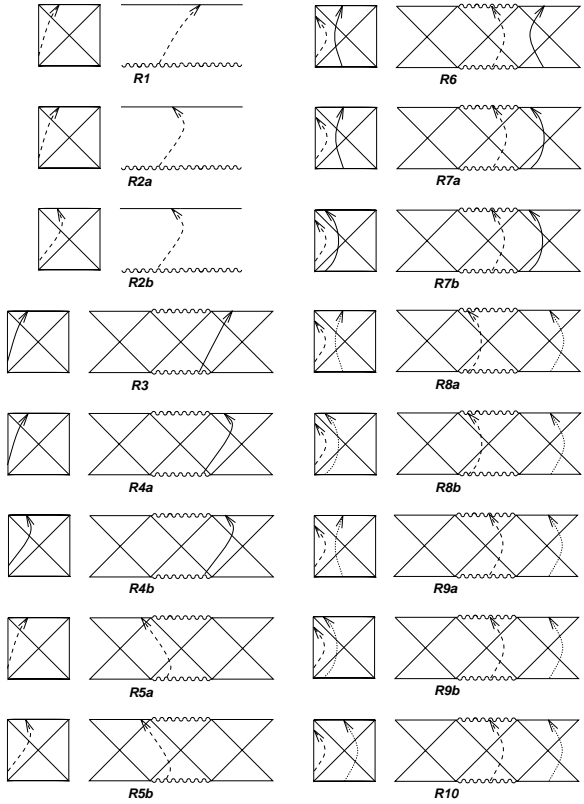}}
\caption{Penrose diagrams representing the motion of the bubble
wall in the interior de~Sitter spacetime (left) and the exterior
Schwarzschild-de~Sitter spacetime (right). The diagrams having the same exterior
but different interior geometries are classified in the same category
with an additional suffix.}
\label{figure3}
\end{figure}

Figure~\ref{figure3} shows the global geometry of the spacetime
represented by Penrose diagrams. In each category, the left side is the
de Sitter spacetime and the right side is the Schwarzschild-de Sitter spacetime.
The dashed or solid curves with arrows denote the trajectories of the
bubble wall as seen by the observers on both sides. The
Schwarzschild-de Sitter spacetime in the first three categories on the left
column does not have event horizons. For examples, the dashed line in
the diagram labelled by $R6$, $R7a$, $R7b$, $R9a$, $R9b$, or $R10$
denotes the formation of a black hole. The dashed line in
the diagram labelled by $R5a$, $R5b$, $R8a$, or $R8b$
denotes the formation of a wormhole.

\begin{figure}
\leavevmode
\hbox{
\epsfxsize=3in
\epsffile{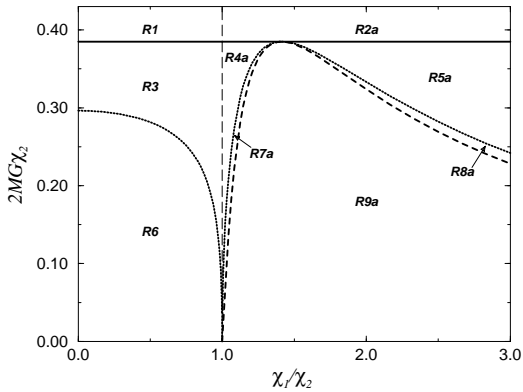}}
\caption{Phase space diagram of the bubble wall dynamics under the constraint
$\kappa^2=|{\chi_1}^2-{\chi_2}^2|$. The boundaries separating
different phases are $M_{\rm ext}$ (solid line),
$M_{\rm max}$ (dotted line), and
$M_{\rm SdS}$ (dashed line).}
\label{figure4}
\end{figure}

\begin{figure}
\leavevmode
\hbox{
\epsfxsize=3in
\epsffile{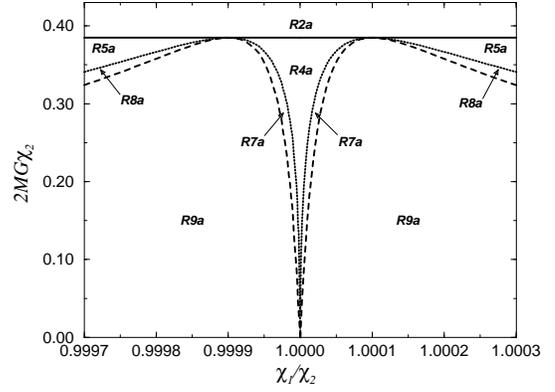}}
\caption{Same as Fig.~\ref{figure4} but for
$\kappa^2=10^4 |{\chi_1}^2-{\chi_2}^2|$.}
\label{figure5}
\end{figure}

\begin{figure}
\leavevmode
\hbox{
\epsfxsize=3in
\epsffile{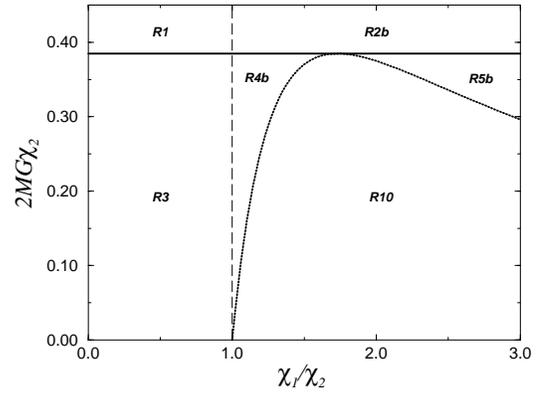}}
\caption{Same as Fig.~\ref{figure4} but for
$\kappa^2=10^{-4} |{\chi_1}^2-{\chi_2}^2|$. The boundaries separating
different phases are $M_{\rm ext}$ (solid line), overlapping with
$M_{\rm max}$ and $M_{\rm SdS}$, and $M_{\rm dS}$ (dotted line).}
\label{figure6}
\end{figure}

We then identify the domain of the dimensionless parameter space
$(2MG\chi_2, \chi_1/\chi_2)$
that fall into the same category, for three representative values of $\kappa$ given by
$\kappa^2=|{\chi_1}^2-{\chi_2}^2|$,
$\kappa^2=10^4 |{\chi_1}^2-{\chi_2}^2|$, and
$\kappa^2=10^{-4} |{\chi_1}^2-{\chi_2}^2|$. Given a set of values of $(2MG\chi_2, \chi_1/\chi_2)$,
both $V(z)$ and $E$ are fixed and the motion of the bubble wall is determined.
The phase-space diagrams are drawn in Fig.~\ref{figure4}, Fig.~\ref{figure5},
and Fig.~\ref{figure6}, respectively, in which the boundaries separating different
phases correspond to the masses defined in Fig.~\ref{figure1}.

\section{Conclusion}\label{conc}

The global structure of the spacetime describing a vacuum bubble in a background
vacuum has been studied using the general-relativistic thin-wall formalism.
In this work, the dynamics of the bubble wall is controlled by the cosmological
constants on both sides and the surface energy density of the wall.
There exists a large parameter space that allows the bubble to collapse
into a black hole or to form a wormhole. The condition for the existence
of this parameter space should be easily satisfied in the context of string landscape.
To investigate the formation rate of black holes and wormholes in the string
landscape is an interesting problem. We may speculate that while slow-roll inflation
is taking place along a flat potential in the string landscape, black holes or wormholes
are being produced by bubble nucleation in quantum tunneling from
the flat plateau to neighboring or even far away local minima with lower vacuum energies
through the landscape barriers, or perhaps by quantum fluctuating to nearby local minima
through shallow barriers. If the formation rate is significant compared to the expansion
rate of the inflation, this may have relevant effects on inflationary physics
and may leave an imprint in the metric perturbation. Overall, this work is
a theoretical study of the dynamics of vacuum bubbles. It itself is an interesting
problem in general relativity, while being an effort towards exploring the observable signatures
of the string landscape scenario.

\begin{acknowledgments}
The authors would like their thanks to H.-C.~Kao, M.~Kleban, S.-Y.~Lin,
Y.~Sekino, and L.~Susskind for helpful discussions.
This work was support in part by the National Science Council,
Taiwan, ROC under the Grants NSC 98-2112-M-001-009-MY3 (KWN)
and NSC 96-2112-M-032-005-MY3 (SYW).
\end{acknowledgments}

\end{document}